\documentclass[a4]{article}

\font\scap=cmcsc10 \hfuzz=5cm
\font\scap=cmcsc10

\font\tenmsb=msbm10
\font\sevenmsb=msbm7
\font\fivemsb=msbm5
\newfam\msbfam
\textfont\msbfam=\tenmsb
\scriptfont\msbfam=\sevenmsb
\scriptscriptfont\msbfam=\fivemsb
\def\Bbb#1{{\fam\msbfam\relax#1}}

\newcount\eqnumber
\def\neweq{{\rm{(\the\eqnumber)}}\global\advance\eqnumber by 1}
\def\eqdef#1{\eqno\xdef#1{\the\eqnumber}\neweq}
\def\newaeq{{\rm{(\the\eqnumber a)}}\global\advance\eqnumber by 1}
\def\eqdaf#1{\eqno\xdef#1{\the\eqnumber}\newaeq}
\def\eqdisp#1{\xdef#1{\the\eqnumber}\neweq}
\def\eqdasp#1{\xdef#1{\the\eqnumber}\newaeq}

\newcount\refnumber
\def\newref{{\the\refnumber}\global\advance\refnumber by 1}
\def\refdef#1{{\xdef#1{\the\refnumber}}\newref}

\begin{document}

\centerline{\bf Calculating the algebraic entropy of mappings with unconfined singularities}
\bigskip
\par\medskip{\scap A. Ramani} and {\scap B. Grammaticos}
\quad{\sl IMNC, Universit\'e Paris VII \& XI, CNRS, UMR 8165, B\^at. 440, 91406 Orsay, France}
\par\medskip{\scap R. Willox} and {\scap T. Mase}\quad
{\sl Graduate School of Mathematical Sciences, the University of Tokyo, 3-8-1 Komaba, Meguro-ku, 153-8914 Tokyo, Japan }
\par\medskip{\scap J. Satsuma}
\quad{\sl  Department of Mathematical Engineering, Musashino University, 3-3-3 Ariake, Koto-ku, 135-8181 Tokyo, Japan}

\bigskip
{\sl Abstract}
\par\smallskip
We present a method for calculating the dynamical degree of a mapping with unconfined singularities. It is based on a method introduced by Halburd for the computation of the growth of the iterates of a rational mapping with confined singularities. In particular, we show through several examples how simple calculations, based on the singularity patterns of the mapping, allow one to obtain the exact value of the dynamical degree for nonintegrable mappings that do not possess the singularity confinement property. We also study linearisable mappings with unconfined singularities to show that in this case our method indeed yields zero algebraic entropy.
\par\bigskip
PACS numbers: 02.30.Ik, 05.45.Yv
\smallskip
Keywords: mappings, integrability, deautonomisation, singularities, algebraic entropy

\bigskip\medskip
1. {\scap Introduction}
\par\medskip
Singularity analysis is the cornerstone of integrability studies. In the words of the late M. D. Kruskal, singularities are ``where interesting things happen''. As is well known, the study of singularities was essential in the case of continuous systems. By requiring that solutions of differential equations be free of multivaluedness-inducing singularities, Painlev\'e was able to obtain integrable second-order differential equations [\refdef\painl] and, in the process, discovered the new transcendents that now bear his name.  Singularities  are equally essential in discrete integrability. The first discrete integrability criterion to be proposed, singularity confinement, was based on the observation that discrete systems that are integrable by spectral methods possess so-called `confined' singularities [\refdef\sincon]. Before proceeding any further, let us clarify the notions of `singularity' and `confinement'. A singularity appears in a  discrete system when at some iteration a degree of freedom is lost. In the case of second-order rational mappings we shall focus on, a loss of a degree of freedom means that the value of $x_{n+1}$ is independent of that of $x_{n-1}$. The term confinement is used to describe the fact that a singularity that appeared at some iteration again disappears after a few more iteration steps, when the lost degree of freedom is recovered by lifting the indeterminacy that arises at that iteration. The existence of unconfined singularities (in the sense that a singularity that appears at some iteration never disappears) typically signals nonintegrability (but it is worth remembering that linearisable systems, generically, possess unconfined singularities [\refdef\tremblay]). However, the confinement property is not a failsafe integrability criterion: there do exist nonintegrable systems that have confined singularities [\refdef\hiv]. 

Confined and unconfined singularities are not the only possible types of singularity. Many systems also possess singularities in which a pattern keeps repeating for all iterations. We have dubbed these singularities `cyclic' [\refdef\rodone]. The existence of cyclic singularities is perfectly compatible with integrability and, as such, cyclic singularities are definitely {\sl not} be to thought of as unconfined. Yet, still another type of singularity does exist, which we refer to as `anticonfined'. These singularities are characterized by patterns in which singular values persist indefinitely in both the forward and backward iteration, with just a finite region of regular values in between. Such patterns can exhibit growth and if the latter is exponential, this is an indication of nonintegrability.

As the singularity confinement property suffered from a lack of sufficiency as a discrete integrability criterion, a different approach was clearly required. To this end, staying within the spirit of singularity analysis, we proposed the method of `full deautonomisation' [\refdef\redemp]. In a nutshell, this method allows one to detect the nonintegrable cases among mappings with confined singularities, by analysing the behaviour of the coefficient functions in a deautonomised version of the mapping that still possesses the same singularity patterns. A different approach, based on the observation that low growth is a characteristic of integrability, was pioneered by Arnold [\refdef\arnold] and Veselov [\refdef\veselov] and was made operational by Bellon and Viallet [\refdef\bellon]. In the approach of the last two authors, adapted to the case of rational mappings, one introduces homogeneous coordinates for the initial values and computes the common degree $d_n$ of the homogeneous polynomials that appear at the $n$th iteration of the mapping. Considering the limit of the degree when $n\to\infty$, one can  introduce the following non-negative quantity
$$\varepsilon=\lim_{n\to\infty}{1\over n}\log d_n,$$
which Bellon and Viallet dubbed `algebraic entropy'. Another useful, related, quantity is the exponential of the algebraic entropy, $\lambda=e^{\varepsilon}$ $(\lambda\geq1)$, which is called the dynamical degree of the mapping. Integrable mappings have a dynamical degree equal to 1 and therefore $\varepsilon=0$. A dynamical degree greater than 1, or equivalently $\varepsilon>0$, indicates nonintegrability. The low growth in integrable mappings results from the fact that the iteration of the mapping induces algebraic simplifications which lower substantially the degree of the iterates. As a result, the degree grows polynomially with the number of iterations, while in the absence of simplifications the degree growth would be exponential. The origin of these simplifications can be traced back to the singularities which appear spontaneously during the iteration of the  mapping and is therefore intimately related to the confinement property.

The value of the algebraic entropy is customarily calculated in a heuristic way, by establishing the behaviour of $d_n$ based on the explicit computation of a sufficient number of iterates. On the other hand, a rigorous approach based on an algebro-geometric analysis of the mapping is also possible [\refdef\favre]. For confining mappings, the latter consists  in performing the regularisation of the mapping, through a (finite) sequence of blow-ups and blow-downs, to an automorphism of the surface obtained from these blow-ups. The dynamical degree of the mapping is then obtained as the largest eigenvalue of this automorphism [\favre], [\refdef\take]. For nonconfining mappings the approach is less systematic, but the general theory  [\favre] tells us that, generically, the dynamical degree will be greater than 1, unless it is a linearisable mapping in which case $\lambda=1$. In fact, in the nonconfining, non-linearisable case, it is known that the dynamical degree is a Pisot number, i.e. a real algebraic integer greater than 1 for which all other roots of its minimal polynomial have modulus less than 1.

Unfortunately, the calculations involved in this algebro-geometric approach are often painstaking and a simpler and faster method was therefore highly needed. An ingeniously simple method was introduced by Halburd in [\refdef\rod]. It was simplified by the present authors into what we call the express method [\rodone].  Halburd's as well as our own method were tailored to the treatment of mappings with confined singularities but both can be extended to the case of unconfined ones. However, the treatment of the latter case is rather delicate when using Halburd's method whereas the application of the express method is much more convenient, whether the mappings with unconfined singularities are integrable (which means, as explained above, that they will be linearisable) or not. The aim of the present paper is to explain this method by means of a selection of illustrative examples.
\par\bigskip
2. {\scap The method of Halburd and the express method: a brief summary}
\par\medskip
In order to calculate the degree growth of the solution of a second-order rational mapping, one starts by introducing a variable, say $z$, through an appropriate choice of initial conditions, $x_0$ and $x_1=z$, where $x_0$ is a completely generic complex number, i.e. it is not supposed to satisfy any particular algebraic relation and  $z\in{\Bbb C}\cup\{\infty\}$. One then calculates the degree of the $n$th iterate of the mapping --  which is now a rational function of $z$, $f_n(z)$ -- as the number of preimages of some arbitrary value $w\in{\Bbb C}\cup\{\infty\}$ for that function. This is tantamount to  counting the  solutions, in $z$, of the equation $f_n(z)=w$ (counted with multiplicity). We would like to point out here that this degree is precisely what Arnold [\arnold] called `complexity'. 

Halburd's method [\rod] consists in computing the degree, not for just any arbitrary value of  $w$, but for special values which appear in the singularity pattern of the mapping. The precise workings of Halburd's method are best explained on an example. Let us consider the mapping
$$x_{n+1}x_{n-1}={(x_n-a)(x_n-b)\over(1-x_n/c)(1-x_n/d)},\eqdef\zena$$
where the $a,b,c,d$ are distinct, non-zero, constants. The confined singularity patterns of (\zena) are
$$\{a,0,b\},\quad\{b,0,a\},\quad\{c,\infty,d\},\quad\{d,\infty,c\}.$$
Moreover, there exist also two cyclic singularity patterns starting from a finite value $x_0$
$$\{\dots x_0,0,x_2,\infty,\dots\}\quad {\rm and}\quad\{\dots x_0,\infty,x_2,0,\dots\},$$
where $x_2$ also takes a finite value. In order to calculate the degree growth of the mapping we compute the number of preimages of the various values which appear in the singularity patterns. We denote by $A,B,C,D$ the number of spontaneous occurrences of $a,b,c,d$. On the singularity patterns we see that the value $a$ can either appear at some iteration step following regular values, or whenever a value $b$ appeared two steps before. We have thus for the degree at iterate $n$, calculated as the number of preimages of the value $a$,  
$$d_n(a)=A_n+B_{n-2}.\eqdef\zdyo$$
Moreover, given that $a$ and $b$ play the same role, we have $B_n=A_n$ and the degree is simply $d_n(a)=A_n+A_{n-2}$. In fact one can easily convince oneself, by just inverting $x$, that $a,b,c$ and $d$ all play the same role, and thus $A_n=B_n=C_n=D_n$. 

Next we turn to the preimages of 0 and $\infty$. Clearly, a 0 appears if it is preceded by $a$ or $b$ in a confined pattern, or one time out of four in each cyclic pattern. The same is true, mutatis mutandis, for $\infty$. We have thus
$$d_n(0)=2A_{n-1}+{1-(-1)^n\over2},\eqdef\ztri$$
(and the same expression for the number of preimages of $\infty$). Equating (\zdyo) and (\ztri) we find
$$A_n-2A_{n-1}+A_{n-2}={1-(-1)^n\over2},\eqdef\ztes$$
the solution of which is
$$A_n=\alpha n+\beta+{n^2\over 4}-{(-1)^n\over8},\eqdef\zpen$$
where the first two terms correspond to the solution of the homogeneous part and the remaining two are due to the source term. Requiring that the degree be equal to 0 and 1 for $n=0$ and $n=1$ respectively, we find the expression
$$d_n={n^2\over2}+{1-(-1)^n\over4}.\eqdef\zhex$$
This is exactly the degree growth obtained empirically: $0, 1, 2, 5, 8, 13, 18, 25, \dots$, which is clearly quadratic. The algebraic entropy is therefore equal to 0, and the mapping integrable.

The express method is far more succinct in nature, as it completely neglects the contribution of the cyclic patterns. In the example above, this means we neglect the source term in equation (\ztes) and instead work with
$$A_n-2A_{n-1}+A_{n-2}=0,\eqdef\zhep$$
the characteristic equation of which is $(\lambda-1)^2=0$. As explained in [\rodone], the absence of a root greater than 1 indicates that the dynamical degree of this mapping is in fact equal to 1 and hence that the mapping should be integrable. The express method thus gives, in a straightforward way, a yes/no answer on the integrable character of the mapping, without having to calculate the exact expression for the degree at each iterate. 
\par\bigskip\goodbreak
3. {\scap Computing the algebraic entropy of nonintegrable, nonconfining mappings}
\par\medskip
As we have shown in [\rodone] on a slew of examples, the express method, if implemented correctly, gives the correct value of the algebraic entropy for both integrable and nonintegrable mappings with confined singularities. In this section we shall show how these results can be extended to nonconfining mappings.

Let us start with 
$$x_{n+1}x_{n-1}={(x_n-a)(x_n-b)\over x_n},\eqdef\zoct$$
where $a,b$ are nonzero, $a\ne b$ and $a^kb^m\ne1$ for any positive integers $k,m$. Two unconfined singularity patterns exist:
$$\{a,0,\infty,\infty^2,\infty,0,ab^2,\infty,\infty,1 /(ab^2),0,\infty,\infty^2,\infty,\dots\}$$
and a similar one starting from $b$. (The notation $\infty^2$ is shorthand for the following: had we introduced a small quantity $\epsilon$ assuming that $x_n=a+\epsilon$ we would have found that $x_{n+1}$ is of the order of $\epsilon$, $x_{n+2}$ of the order of $1/\epsilon$ and $x_{n+3}$ of the order of $1/\epsilon^2$). Cyclic patterns do also exist for (\zoct): $\{x_0,0,\infty,\infty^2,\infty,0\}$ and $\{x_0,\infty,\infty\}$. 

We denote the number of spontaneous occurrences of $a,b$ by $A,B$ respectively and we shall use these quantities to calculate the dynamical degree of the mapping by an express ansatz, i.e. by expressing the number of preimages of the values that appear in the singularity patterns but {\sl neglecting} any contributions of possible cyclic patterns, in the spirit of the express method for confining mapings as explained in section 2. We then have $d_n(a)=A_n=B_n$ for the number of preimages of $a$ or $b$, since $b$ plays exactly the same role as $a$, and for the preimages of 0 we write
$$d_n(0)=(A+B)_{n-1}+(A+B)_{n-5}+(A+B)_{n-10}+(A+B)_{n-14}+\dots=2\sum_{k=0}^{\infty}(A_{n-9k-1}+A_{n-9k-5}),\eqdef\zenn$$
where (given our assumed initial conditions, $x_0$ constant and $x_1=z$) we use the convention that $A_m$ and $B_m$ are always zero for $m\leq0$.
Similarly, for the preimages of $\infty$ we have
$$\displaylines{d_n(\infty)=(A+B)_{n-2}+2(A+B)_{n-3}+(A+B)_{n-4}+(A+B)_{n-7}+(A+B)_{n-8}+\dots\hfill\cr\hfill=2\sum_{k=0}^{\infty}(A_{n-9k-2}+2A_{n-9k-3}+A_{n-9k-4}+A_{n-9k-7}+A_{n-9k-8}).\quad\eqdisp\zdek\cr}$$
Equating the expressions for $d_n(0)$ and $d_n(\infty)$ 
we obtain a finite linear (homogeneous) recursion relation for $A_n$ in which we substitute $A_n=\lambda^n$ in order to find the roots of its characteristic equation. At this point we assume that this characteristic equation has a root with modulus greater than 1 and we take the limit $n\to\infty$. This allows us to sum the infinite series that appear, thus obtaining the relation
$${2\over1-{1\over\lambda^9}}\left({1\over\lambda}+{1\over\lambda^5}\right)={2\over1-{1\over\lambda^9}}\left({1\over\lambda^2}+{2\over\lambda^3}+{1\over\lambda^4}+{1\over\lambda^7}+{1\over\lambda^8}\right),\eqdef\dena$$
or
$$\lambda^7-\lambda^6-2\lambda^5-\lambda^4+\lambda^3-\lambda-1=0,\eqdef\denb$$
the largest root of which is also that of (\dena).
Equation (\denb) defines the Pisot number $\lambda=2.10455\dots$, in very good agreement with the dynamical degree calculated from the sequence of homogeneous degrees that is obtained empirically from (\zoct): 0, 1, 2, 4, 9, 19, 40, 84, 177, 372, 783, 1648, \dots. 

In the computation of the dynamical degree above we have used the expressions for the degree obtained from $d_n(0)$ and $d_n(\infty)$. However a third expression does exist, obtained from $d_n(a)$, and it is interesting to verify that the conclusions one reaches working with the latter are indeed consistent with the previous results. We have for instance, equating $d_n(a)$ to $d_n(0)$, the relation
$$1={2\over1-{1\over\lambda^9}}\left({1\over\lambda}+{1\over\lambda^5}\right), \eqdef\RWI$$
which leads back to (\denb) up to an inconsequential factor $(\lambda^2-\lambda+1)$.

Obviously our method works for additive mappings as well. For example, the mapping
$$x_{n+1}+x_{n-1}=a+{1\over x_n^2},\eqdef\ddyo$$
has the unconfined singularity pattern
$$\{0,\infty^2,a,\infty^2,0,\infty^2,a,\dots\}.$$
This mapping is not integrable as can be assessed from the sequence of homogeneous degrees: 0, 1, 2, 5, 12, 29, 68, 161, 380, 897, 2116, $\dots$, which exhibits exponential growth. The dynamical degree can be easily computed following the method we introduced just above. We denote by $Z_n$ the spontaneous occurrences of 0 and  write for the preimages of 0 the expression
$$d_n(0)=Z_n+Z_{n-4}+Z_{n-8}+\dots=\sum_{k=0}^{\infty}Z_{n-4k},\eqdef\dtri$$
where, again, $Z_m=0$ whenever $m\leq0$.
Similarly, from the preimages of infinity we obtain the expression
$$d_n(\infty)=2Z_{n-1}+2Z_{n-3}+2Z_{n-5}+\dots= 2 \sum_{k=0}^{\infty}Z_{n-2k-1}.\eqdef\dtes$$
By equating these two expressions we obtain a linear recursion for $Z_n$ in which we substitute $Z_n=\lambda^n$ while taking $n\to\infty$, assuming that the associated characteristic equation has a root greater than 1. We thus find
$${1\over1-{1\over\lambda^4}}={2\over\lambda}\,{1\over1-{1\over\lambda^2}},\eqdef\dpen$$
or, 
$$\lambda^3-2\lambda^2-2=0.\eqdef\dhex$$
The largest root of the latter is the Pisot number $\lambda=2.3593\dots$, in very good agreement with the dynamical degree obtained from the degree growth in the sequence of degrees given above.
\par\bigskip
4. {\scap The case of linearisable mappings}
\par\medskip
As we pointed out in the introduction, linearisable mappings do not, in general, possess the confinement property. Their growth is in fact slower than that of mappings integrable through spectral methods, the degree growing linearly (or even saturating)  with the number of iterations, a property that constitutes a linearisability criterion. As we have explained in previous publications, second-order linearisable mappings come in three varieties: projective, Gambier type, and a type of mapping that we have dubbed third kind. We refer to [\refdef\linear] for more details concerning these three classes. It turns out that projective mappings do not present any interest for the present study and we shall concentrate on the Gambier and third-kind ones. 

We start with the Gambier mapping
 $$x_{n+1}=ax_{n-1}{x_n-a\over x_n-1},\eqdef\dhep$$
with $a\neq 0, 1$ and $a^{2k+1}\neq 1$ for any positive integer $k$. 
This mapping was derived in [\refdef\tsuda], where we showed that it can be written as two homographic mappings in cascade: $y_n=ay_{n-1}$ and $x_{n+1}=a(x_n+y_n)/(x_n-1)$. Its homogeneous degree growth can be easily calculated:  0, 1, 1, 2, 2, 3, 3, 4, 4, 5, 5, 6, 6, \dots. Two singularity patterns exist for (\dhep), a confined one
$$\{1,\infty,a\},$$
and an unconfined one
$$\{a,0,a^3,0,a^5,0,\dots\}.$$
We denote by $A$ and $U$ the spontaneous occurrences of $a$ and 1. Under our express ansatz we have that the number of preimages of the value 1 is just $U_n,$ while that of $a$ is $A_n+U_{n-2}$. For the preimages of the value 0 we have
$$d_n(0)=A_{n-1}+A_{n-3}+A_{n-5}+\dots=\sum_{k=0}^{\infty}A_{n-2k-1}.\eqdef\RWII$$
Equating these expressions we obtain $A_n=U_n-U_{n-2}$ and finally a linear recursion relation for $U_n$. In order to obtain the characteristic equation of the latter we put $U_n=\lambda^n$ and take the limit $n\to\infty$, assuming the existence of a root $\lambda>1$. We find
$${1\over\lambda} {1\over 1-1/\lambda^2} (1-1/\lambda^2)= 1,\eqdef\RWIII$$ 
with solution $\lambda=1$. This means that the root $\lambda > 1$ that we postulated cannot exist. However, as by definition the dynamical degree cannot be less than 1, this shows that the dynamical degree of (\dhep) must be equal to 1, which   is in perfect agreement with the integrable, and indeed linearisable, character exhibited by the degree growth for this mapping,  given above.

The second mapping of Gambier type we are going to study is
$${x_{n+1}+x_n\over x_{n-1}+x_n}={1-x_{n}\over 1+x_{n}},\eqdef\doct$$
which, as was shown in [\refdef\tsudanon], can be written as the system $y_n=y_{n-1}+1$, $x_{n+1}=(1-x_ny_n)/(y_n-1)$. Equation (\doct) has two singularity patterns, the second being unconfined:
$$\{1,-1\}\quad{\rm and}\quad\{-1,\infty,\infty,\infty,\dots\},$$
and the growth of the homogeneous degrees is perfectly linear: 0, 1, 2, 3, 4, 5, 6, 7, 8, 9, \dots. Moreover an anticonfined singularity, i.e. one in which a finite number of regular values are bracketed by singularities, does also exist: 
$$\{\dots,\infty,\infty,\infty,x,\infty, \infty,\infty,\dots\}.$$
The essence of the express method is to neglect also the contribution of the anticonfined singularity (with one caveat that will be addressed in section 5). 

We denote by $U$ and $M$ the spontaneous occurrences of $+1$ and $-1$, respectively. From the preimages of the values $+1$ and $-1$ we find that $d_n(1)=U_n$ and $d_n(-1)=M_n+U_{n-1}$. On the other hand, for the preimages of the value $\infty$ we have
$$d_n(\infty)=M_{n-1}+M_{n-2}+M_{n-3}+\dots=\sum_{k=1}^{\infty}M_{n-k}.\eqdef\RWIV$$
Equating the three expressions for the degree we obtain
$$U_n=M_n+U_{n-1}=\sum_{k=1}^{\infty}M_{n-k}.\eqdef\RWV$$
We can remark here that the contribution of the value $\infty$ of the anticonfined pattern would have generated a constant non-homogeneous term in the linear equation above, just as in the case of cyclic patterns, having no effect on the roots of the characteristic equation.

Next we substitute $U_n=U_0\lambda^n$ and $M_n=M_0\lambda^n$, while still assuming there is a characteristic root greater than 1 in order to be able to  resum the geometric series that arise in the limit $n\to\infty$. We obtain finally
$$U_0={M_0\over\lambda-1}\quad{\rm and}\quad M_0=U_0\left(1-{1\over\lambda}\right),\eqdef\RWVI$$
and find $\lambda=1$ as the only solution. 
This excludes $\lambda>1$  as a possibility and by the same argument as for mapping (\dhep) we conclude that $\lambda=1$ is the value of the dynamical degree. This is indeed the expected value for the dynamical degree of (\doct), given the integrable character of the latter.

Next we turn to a mapping of what we call the third-kind type. These mappings (at least all the known ones) have the distinctive feature that at the autonomous limit they do not have unconfined singularities. However, when deautonomised they do possess such singularities, and thus we can apply to them the same treatment as in the preceding examples.
We start with the autonomous mapping
$${\zeta\over x_n+x_{n+1}}+{\zeta\over x_n+x_{n-1}}=1+{z\over x_n},\eqdef\denn$$
which is {\sl not} integrable in general, let alone linearisable of the  third-kind. It is easy to show that (\denn) has two unconfined singularities $\{0,0,0,\dots\}$ and $\{\infty,\infty,\infty,\dots\}$. However when $\zeta$ and $z$ are related through $\zeta=z$ or $\zeta=2z$ the mapping does become linearisable of the third kind [\refdef\angers]. In what follows we shall focus on these two linearisable cases.

We start with the case $\zeta=2z$: 
$${2z\over x_n+x_{n+1}}+{2z\over x_n+x_{n-1}}=1+{z\over x_n}.\eqdef\ddek$$
This mapping  has a single confined singularity  $\{0,0\}$ but it also possesses an anticonfined one:
$$\{\dots,\infty^2,\infty,\infty,x,x',x'',\infty, \infty,\infty^2, \infty^2,\infty^3,\infty^3,\dots\}.$$
Contrary to the case of the Gambier mapping (\doct) this anticonfined pattern exhibits growth in the orders of the singularity. The deautonomisation of (\ddek) was presented in [\linear], together with its explicit integration. We found
$${z_n+z_{n+1}\over x_n+x_{n+1}}+{z_n+z_{n-1}\over x_n+x_{n-1}}=1+{z_n\over x_n},\eqdef\vena$$
where $z_n$ is a free function of the indepedent variable. In general, the deautonomisation alters the singularity patterns for third-kind mappings. While the confined pattern is preserved, the anticonfined one becomes the unconfined pattern $\{\infty,\infty,\infty,\dots\}$. (Note that all infinities have exponent 1). We can now apply the method developed in the previous sections to compute the dynamical degree of (\vena). Unfortunately, from these two singularity patterns, we do not obtain a number of equations sufficient for the application of the method. In order to overcome this difficulty we apply the method we developed in [\rodone] and which consists in introducing an auxiliary variable. We define $y_n=(z_n+z_{n+1})/(x_n+x_{n+1})$ and rewrite equation (\vena) as
$$x_n+x_{n+1}={z_n+z_{n+1}\over y_n}\eqdaf\vdyo$$
$$y_n+y_{n-1}=1+{z_n\over x_n}.\eqno(\vdyo b)$$
The singularity patterns are now $\{x=0,y=\infty,x=0\}$ and $\{y=0,x=\infty,y=1,x=\infty,y=0,x=\infty, y=1, x=\infty,y=0,\dots\}$. We introduce the notations $X$ and $Y$ for the number of spontaneous occurrences of $x=0$ and $y=0$. We have for the number of preimages of $x=0$ and $y=\infty$ respectively,
$$d_n(x=0)=X_n+X_{n-1}\quad{\rm and}\quad d_n(y=\infty)=X_{n}.\eqdef\RWVII$$
Similarly, for the preimages of $x=\infty$ and $y=0$ we have
$$d_n(x=\infty)=Y_{n-1}+Y_{n-2}+\dots=\sum_{k=0}^{\infty}Y_{n-k-1}\quad{\rm and}\quad d_n(y=0)=Y_{n}+Y_{n-2}+\dots=\sum_{k=0}^{\infty}Y_{n-2k}.\eqdef\RWVIII$$ 
In order to obtain the corresponding characteristic equation we put $X_n=X_0\lambda^n$ and $Y_n=Y_0\lambda^n$ and look for the root with the largest modulus, assuming that the latter is greater than 1 so that we can resum the geometric series that arise when we let $n$ tend to $\infty$. Equating the expressions for the preimages of the respective values of $x$, as well as for $y$,  obtained above, we find
$$X_0\left(1+{1\over\lambda}\right)=Y_0{1\over\lambda}\left({1\over1-{1\over\lambda}}\right)\eqdef\RWIX$$
and 
$$X_0=Y_0\left({1\over1-{1\over\lambda^2}}\right).\eqdef\RWX$$
Combining these two equations we arrive at $\lambda=1$ which excludes $\lambda>1 $  as a possibility and we conclude that $\lambda=1$ is indeed the value of the dynamical degree.

The case $\zeta=z$ can be treated along similar lines. The mapping in its non-autonomous form is [\angers]
$${z_n+z_{n+1}\over x_n+x_{n+1}}+{z_n+z_{n-1}\over x_n+x_{n-1}}=1+{z_{n+1}+z_{n-1}\over x_n},\eqdef\vtri$$
or equivalently,
$$x_n+x_{n+1}={z_n+z_{n+1}\over y_n}\eqdaf\vtes$$
$$y_n+y_{n-1}=1+{z_{n+1}+z_{n-1}\over x_n}.\eqno(\vtes b)$$
The singularity patterns are now $\{y=0,x=\infty,y=1,x=\infty,y=0\}$ and $\{x=0,y=\infty,x=0,y=\infty,x=0,y=\infty,\dots\}$. Again we introduce $X$ and $Y$ for the number of spontaneous occurrences of $x=0$ and $y=0$, allowing us to write for the number of preimages of $y=0$ and $x=\infty$,
$$d_n(y=0)=Y_n+Y_{n-2}\quad{\rm and}\quad d_n(x=\infty)=Y_{n-1}+Y_{n-2}.\eqdef\RWXI$$
Similarly, for the preimages of $x=0$ and $y=\infty$ we have
$$d_n(x=0)=X_{n}+X_{n-1}+\dots=\sum_{k=0}^{\infty}X_{n-k}\quad{\rm and}\quad d_n(y=\infty)=X_{n}+X_{n-1}+\dots=\sum_{k=0}^{\infty}X_{n-k}.\eqdef\RWXII$$
Again we look for the characteristic equation putting $X_n=X_0\lambda^n$ and $Y_n=Y_0\lambda^n$ and with the assumption $\lambda>1$ for the largest root, we find for the degrees of $x$ and $y$
$$Y_0{1\over\lambda}\left(1+{1\over\lambda}\right)=X_0\left({1\over1-{1\over\lambda}}\right)\eqdef\RWXIII$$
and 
$$Y_0\left(1+{1\over\lambda^2}\right)=X_0\left({1\over1-{1\over\lambda}}\right).\eqdef\RWXIV$$
Combining the two equations we find again $\lambda=1$ and conclude that the dynamical degree is indeed 1.

At this point one may wonder what happens in the generic case (\denn). Again we introduce the auxiliary variable $y_n=\zeta/(x_n+x_{n+1})$. The singularity patterns are now $\{y=0,x=\infty,y=1,x=\infty,y=0,x=\infty,y=1,x=\infty\}$ and $\{x=0,y=\infty,x=0,y=\infty,x=0,y=\infty,\dots\}$. Denoting by $X$
 and $Y$  the number of spontaneous occurrences of $x=0$ and $y=0$, respectively, we find the equations
$$\sum_{k=0}^{\infty}Y_{n-k-1}=\sum_{k=0}^{\infty}X_{n-k},\eqdef\RWXVI$$
$$\sum_{k=0}^{\infty}Y_{n-2k}=\sum_{k=0}^{\infty}X_{n-k}.\eqdef\RWXV$$
With the usual assumption on the largest characteristic root we find 
$${Y_0\over \lambda(1-{1\over\lambda})}={X_0\over 1-{1\over\lambda}}={Y_0\over 1-{1\over\lambda^2}},\eqdef\ARI$$
and, finally, that the dynamical degree should satisfy the equation $\lambda^2-\lambda-1=0$, i.e. that it is equal to the golden mean $\varphi=(1+\sqrt 5)/2$. This should correspond to the degree growth ratio of $x$ and $y$. Computing the degree of the iterates of equation (\denn) for $x$ we find the sequence 0, 1, 3, 6, 11, 19, 32, 53, 87, 142, 231, 375, 608, 985, 1595, \dots, in excellent agreement with the value of $\lambda$ just obtained.
\par\bigskip
5. {\scap The effect of anticonfinement}
\par\medskip
In the previous section we encountered mappings with anticonfined singularities, where the effect of the anticonfined part was inconsequential, at least in the spirit of the express method. However those mappings were special, in that they are linearisable. It thus does make sense to examine the effect of anticonfined singularities in the case of non-linearisable mappings. 

The first example is a mapping that was first proposed in [\tsuda] and studied in [\redemp,\refdef\anticonf]:
$$x_{n+1}=x_{n-1}\left( x_n-{a_n^2\over x_n}\right).\eqdef\tena$$
As shown in [\redemp], (\tena) has two confined singularities when $a_n$ satisfies the constraint $a_{n+2}=a_n^2a_{n-1}$. Here we shall assume that this condition is {\sl not} satisfied. We obtain thus the sequence of degrees: 0, 1, 2, 4, 8, 16, 32, 64,\dots. In this case the mapping has the following unconfined singularities
$$\{\pm a_n,0,\infty,\mp a_{n+1}a_{n+2}^2, \infty,\infty,\infty^2,\infty^3,\infty^5,\cdots\},$$
the exponents of $\infty$ following a Fibonacci sequence. However, as shown in [\anticonf], (\tena) also has an anticonfined singularity of the form
$$\{\cdots,0^5,0^3,0^2,0,0,x,0,\infty,x', \infty,\infty,\infty^2,\infty^3,\infty^5,\cdots\},$$
which, by the way, also exhibits a Fibonacci sequence in the orders of $\infty$.

In order to compute the degree growth we denote by $A_n$ the number of spontaneous occurrences of $\pm a_n$. Obviously, given the form of (\tena), $+a_n$ and $-a_n$ play the same role and we have thus
$$d_n(a_n)=d_n(-a_n)=A_n.\eqdef\RWXVII$$
For the preimages of 0, neglecting the occurrence of 0 in the anticonfined pattern, we have simply 
$$d_n(0)=2A_{n-1},\eqdef\RWXVIII$$
(the factor 2 being due to the contribution  of  $+a_n$ and $-a_n$). Equating the two expressions for the degree of the mapping we obtain of course $\lambda=2$ for the dynamical degree, in perfect agreement with the computed growth. 

At this point it is interesting to look at the effect of the preimages of $\infty$ (neglecting any contribution from the anticonfined pattern). We have
$$d_n(\infty)=2(A_{n-2}+A_{n-4}+A_{n-5}+2A_{n-6}+3A_{n-7}+5A_{n-8}+\cdots).\eqdef\RWXIX$$
Equating the two expressions for $d_n(0)$ and $d_n(\infty)$ we find
$$A_{n-1}=A_{n-2}+\sum_{k=1}^{\infty}f_kA_{n-k-3},\eqdef\RWXX$$
where the $f_n$ obey the Fibonacci recursion $f_k=f_{k-1}+f_{k-2}$ with $f_0=0$ and $f_1=1$. In order to obtain the roots of the characteristic equation we put $A_n\propto\lambda^n$, assuming the existence of a root $\lambda>1$, and find readily
$$1={1\over\lambda}+{S\over\lambda^2}\eqdef\RWXXI$$
where $S=\sum_{k=1}^{\infty}f_n\lambda^{-k}$. The computation of $S$ is performed assuming that the series converges, i.e. that the $\lambda^k$ term in the denominator grows faster than the Fibonacci number $f_k\propto \varphi^k$ in the numerator, where $\varphi$ represents the golden mean. With this assumption we obtain $S=\lambda/(\lambda^2-\lambda-1)$ whereupon we find that the largest root of the characteristic equation is $\lambda=2$, confirming the value of the dynamical degree obtained above. Since this value of $\lambda$ is indeed larger than the golden mean, our assumption of convergence for the series $S$ was justified.

In the example above we have implemented the express method neglecting any contribution coming from the anticonfining pattern.  It is therefore interesting to verify that this contribution is indeed subdominant even though it is exponentially growing. Since we have two free values of $x$ in the anticonfined pattern we have two instances of sequences of infinities, resulting in a global contribution $F_n=\delta_{0,n-2}+f_{n-3}+f_n$, for $n\ge 1$, growing like $\varphi^k$, which is indeed negligible compared to $2^k$.

Next we turn to the mapping 
$$x_{n+1}={a x_{n-1}\over (1-x_n^q)},\eqdef\ttri$$
where $q$ is a positive integer and $a$ is a non-zero constant such that $a^m\ne 1$ for any integer $m$. The mapping possesses $q$ unconfined singularities
$$\{h,\infty,0^q\infty,0^{2q},\infty,0^{3q},\infty,0^{4q},\cdots\},$$
one for each solution of $h^q=1$. A cyclic singularity does also exist $\{x,0\}$, as well as an anticonfined one
$$\{\cdots,\infty^{g_5},\infty^{g_4}, \infty^{g_3},\infty^{g_2},\infty^{g_1},x,\infty,0^q\infty,0^{2q},\infty,0^{3q},\infty,0^{4q},\cdots\},$$
where the exponents $g_n$ of the infinities obey the recursion relation $g_n=qg_{n-1}+g_{n-2}$ with $g_0=0$ and $g_1=1$.

We denote by $H_n$ the number of spontaneous occurrences of $h$, from the number of preimages of which we find the degree
$$d_n(h)=H_n,\eqdef\RWXXII$$
for each root $h$ of the equation $h^q=1$. Summing the contributions coming from these $q$ roots and neglecting the contributions in the anticonfined pattern, we have for the preimages of $\infty$
$$d_n(\infty)=q(H_{n-1}+H_{n-3}+H_{n-5}+\cdots)=q\sum_{k=0}^{\infty}H_{n-2k-1}.\eqdef\RWXXIII$$
 In order to obtain the roots of the characteristic equation for the relation $d_n(h)=d_n(\infty)$, we put $H_n\propto\lambda^n$ in (\RWXXII) and (\RWXXIII), assuming there exists a root $\lambda>1$, and find 
$$1={q\lambda\over\lambda^2-1},\eqdef\RWXXIV$$
which yields $\lambda=(q+\sqrt{q^2+4})/2$ as the dynamical degree for this mapping. 
The same result can of course also be obtained using the preimages of 0, although the calculation is slightly more involved. Neglecting the  contribution from the anticonfined pattern, we have 
$$d_n(0)=q(qH_{n-2}+2qH_{n-4}+3qH_{n-6}+\cdots),\eqdef\RWXXV$$
leading to the equation
$$H_n=q^2\sum_{k=1}^{\infty}kH_{n-2k}.\eqdef\RWXXVI$$
The characteristic equation becomes now
$$1=\left({q \lambda\over\lambda^2-1}\right)^2,\eqdef\RWXXVII$$
which, since $\lambda$ is assumed to be greater than 1, again leads to the same equation for the dynamical degree $\lambda^2-q\lambda-1=0$. Note that, had we taken into account the contributions of the cyclic and anticonfined patterns, we would have had to include a source term in (\RWXXVI), consisting of a constant term 1 (due to the cyclic pattern) and of a linearly increasing term due to the  linear growth of the orders of $0$ every two steps in the anticonfined pattern. Although unbounded, as this growth is much slower than the exponential one coming from the homogeneous part of the equation, it is the latter growth that dominates and the value of the dynamical degree obtained from this more detailed calculation is therefore unchanged from the one we obtained, more easily, by the express method. 

It is interesting at this stage to consider the backward evolution of the mapping 
$$x_{n-1}={x_{n+1}\over a} (1-x_n^q).\eqdef\ttri$$
As expected $q$ unconfined singularities exist
$$\{\cdots,0,{h\over a^3},0,{h\over a^2},0,{h\over a},0,h\},$$
one for each solution of $h^q=1$. The anticonfined and the cyclic singularities are obviously the same as for the forward evolution.
With the same notations as for the forward evolution we find for the degree
$$d_n(h)=H_n.\eqdef\RWXXVIII$$
Similarly, for the preimages of $0$ we have
$$d_n(0)=q(H_{n-1}+H_{n-3}+H_{n-5}+\cdots).\eqdef\RWXXIX$$
This obviously leads to the same characteristic equation as for the forward evolution, $\lambda^2-q\lambda-1=0$. 
Note, however, that one would have obtained the very same equation from the recursion relation for the exponents of the infinities in the anticonfined pattern for the backward map. In fact, whereas the growth obtained from an anticonfined pattern, in general, constitutes a lower bound for the overall growth of the mapping, this  lower bound here turns out to be exactly equal to the dynamical degree of the mapping, contrary to the case of the forward mapping where the anticonfined pattern indeed only yielded a lower bound. Moreover, as no infinities appear in the unconfined singularity pattern for the backward mapping (\ttri), the contribution of the anticonfined one completely determines $d_n(\infty)$ but does not intervene at all in the calculation involving $d_n(0)$ and $d_n(h)$. This leads us to a small but important caveat: when using the express method to obtain the dynamical degree for a mapping with anticonfined singularities that exhibit exponential growth in the orders of the singularities, one should always check whether this growth is subdominant compared to the characteristic exponent obtained from the express method. If this is the case, the dynamical degree is given by this exponent. If not, the degree growth of the mapping is at least as fast as the fastest growth in the anticonfined pattern. In both cases the mapping is guaranteed to be nonintegrable.
\par\bigskip
6. {\scap Conclusion}
\par\medskip
Computing the degree growth of a rational mapping and deducing its dynamical degree (or, equivalently, its algebraic entropy) is an arduous task. The empirical approach consisted in computing explicitly a sequence of degrees and establishing a recursion relation for the latter. (Admittedly, this calls for some experience but it is not an unmanageable task). The rigorous method consisted in performing the full algebro-geometric analysis of the  mapping and deducing from it  the dynamical degree. The advantage of the latter method is that it provides an exact result but the price to pay are delicate and sometimes quite lengthy calculations. 

Clearly, this situation called for a different approach. A first step in this direction was the introduction of the full deautonomisation method which allows one to compute exactly the algebraic entropy of nonintegrable mappings with confined singularities [\redemp]. However, a major accomplishment was the proposal by Halburd [\rod] of an, in essence, elementary method which makes it possible to obtain the exact degrees of the iterates of mappings with confined singularities, be they integrable or not. The power of the method resides in the proper use of the singularity structure of the mapping at hand. Based on the fact that the degree can be obtained from the number of preimages of some generic value, Halburd's method uses the values appearing in the singularity pattern in order to establish recursion relations leading to the degree. The method necessitates the precise knowledge of all singularity patterns, including cyclic ones. However, as we showed in our express method [\rodone], knowledge of the cyclic patterns is not necessary in order to assess the integrability of a confining mapping.

In this paper we extended our approach to the treatment of mappings with unconfined singularities. The essence of our approach is to establish -- a priori finite -- recursion relations based on the unconfined singularities, and to analyse the limit in which the sums in these recursion relations become infinite series. We do this by assuming that the largest root of the corresponding characteristic equation is greater than 1, which allows us to resum all the infinite series that appear. This largest root then yields the value of the dynamical degree of the mapping. Interestingly enough, our method is also applicable to linearisable mappings which (may) have unconfined singularities, but still with zero algebraic entropy. The trick consists in resumming the infinite series that appear by assuming that the largest characteristic root is still greater than 1. This then leads to a contradiction showing that, as expected, the dynamical degree has to be 1 after all, 

We conclude that our approach is yet another proof (if there is still a need for one) of the importance of singularity analysis in the study of integrability.

\bigskip
{\scap Acknowledgements}
\par\medskip
RW would like to acknowledge support from the Japan Society for the Promotion of Science (JSPS),  through the JSPS grant: KAKENHI grant number 15K04893. TM would also like to acknowledge support from JSPS through the grant 16H06711 and JS through the grant 16K05048.

\par\bigskip
{\scap References}
\begin{description}
\item{[\painl]} P. Painlev\'e, Acta Math. 25 (1902) 1.
\item{[\sincon]} B. Grammaticos, A. Ramani and V. Papageorgiou, Phys. Rev. Lett. 67 (1991) 1825.
\item{[\tremblay]} A. Ramani, B. Grammaticos and S. Tremblay, J. Phys. A 33 (2000) 3045
\item{[\hiv]} J Hietarinta and C-M. Viallet, Phys. Rev. Lett. 81, (1998) 325.
\item{[\rodone]} A. Ramani, B. Grammaticos, R. Willox and T. Mase, J. Phys. A 50 (2017) 185203.
\item{[\redemp]} A. Ramani, B. Grammaticos, R. Willox, T. Mase and M. Kanki, J. Phys. A 48 (2015) 11FT02.
\item{[\arnold]} V. I. Arnold, Bol. Soc. Bras. Mat. 21 (1990) 1.
\item{[\veselov]} A.P. Veselov, Comm. Math. Phys. 145 (1992) 181.
\item{[\bellon]} M. Bellon and C-M. Viallet, Comm. Math. Phys. 204 (1999) 425.
\item{[\favre]} J. Diller and C. Favre, Amer. J. Math. 123 (2001) 1135.
\item{[\take]} T. Takenawa, J. Phys. A 34 (2001) 10533.
\item{[\rod]} R.G. Halburd, Proc. R. Soc. A 473 (2017) 20160831.
\item{[\linear]}  A. Ramani, B. Grammaticos and J. Satsuma, J. Phys. A 45 (2012) 365202.
\item{[\tsuda]}  T. Tsuda, A. Ramani, B. Grammaticos and T. Takenawa, Lett. Math. Phys. 82 (2007) 39.
\item{[\tsudanon]}  A.S. Carstea, A. Ramani and B. Grammaticos, J. Phys. A 42 (2009) 485207.
\item{[\angers]} A. Ramani, B. Grammaticos and T. Tamizhmani,  S\'eminaires et Congr\`es de la SMF 14 (2006) 23.
\item{[\anticonf]}  T. Mase, R. Willox, B. Grammaticos and A. Ramani, {\sl Integrable mappings and the notion of anticonfinement}, preprint (2017) arXiv:1511.02000v2 [math-ph].
\end{description}

\end{document}